\documentclass[twocolumn]{autart}

\usepackage{amsfonts, amsmath, amstext, amssymb}
\usepackage{graphicx}
\usepackage{algorithm}

\usepackage{algcompatible}
\usepackage[dvipsnames]{xcolor}
\usepackage{cite}
\usepackage{accents}
\usepackage{acronym}
\usepackage{subcaption}
\usepackage{paralist}
\usepackage{tikz}
\usepackage[a-1b]{pdfx}
\usetikzlibrary{shapes,arrows,positioning,fit,calc}
\usepackage{bookmark}
\newtheorem{lemma}{Lemma}
\newtheorem{theorem}{Theorem}
\newtheorem{corollary}{Corollary}
\newtheorem{proposition}{Proposition}

\newtheorem{assumption}{Assumption}

\newcommand{\reals}{\mathbb{R}}
\newcommand{\naturals}{\mathbb{N}}
\newcommand{\X}{\mathcal{X}}
\newcommand{\R}{\mathcal{R}}
\newcommand{\D}{\mathcal{D}}
\newcommand{\C}{\mathcal{C}}
\newcommand{\Cinf}{\mathcal{C}_\infty}
\renewcommand{\O}{\mathcal{O}}
\newcommand{\Oinf}{\mathcal{O}_\infty}

\newcommand{\trans}{{\top\!}}

\renewcommand{\qed}{\hfill\blacksquare}

\acrodef{mpc}[\sc mpc]{model predictive control}
\acrodef{edmd}[\sc edmd]{extended dynamic mode decomposition}
\acrodef{siso}[\sc siso]{single-input single-output}
\acrodef{iss}[{\sc iss}]{input-to-state stable}
\acrodef{ci}[{\sc ci}]{controlled invariant}
\acrodef{mci}[{\sc mci}]{maximal \ac{ci}}
\acrodef{pi}[{\sc pi}]{positive invariant}
\acrodef{mpi}[{\sc mpi}]{max-\ac{pi}}
\acrodef{sdp}[{\sc sdp}]{semi-definite program}
\acrodef{sos}[{\sc sos}]{sum of squares}
\acrodef{lp}[{\sc lp}]{linear program}
\acrodef{ip}[{\sc ip}]{integer program}
\acrodef{ls}[{\sc ls}]{least-squares}
\acrodef{cg}[{\sc cg}]{command governor}
\acrodef{pe}[{\sc pe}]{persistently exciting}
\acrodef{lti}[{\sc lti}]{linear time-invariant}
\acrodef{lqr}[{\sc lqr}]{linear–quadratic regulator}
\acrodef{pid}[{\sc pid}]{proportional-integral-differential}

\begin{document}

\begin{frontmatter}

\title{Data-Driven Invariant Set for Nonlinear Systems with application to Command Governors}

\thanks[footnoteinfo]{This material is based upon work supported by the National Science Foundation under NSF Grants CMMI-2105631 and CMMI-2303157. Any opinions, findings, and conclusions or recommendations expressed in this material are those of the authors and do not necessarily reflect the views of the National Science Foundation.
}
\author[Paestum]{Ali Kashani}\ead{kashani@unm.edu},               
\author[Paestum]{Claus Danielson}\ead{cdanielson@unm.edu}  

\address[Paestum]{Department of Mechanical Engineering, University of New Mexico, Albuquerque, NM, US 87106}  

\vspace{-2em}
\begin{abstract}
This paper presents a novel approach {to synthesize} positive invariant sets for unmodeled nonlinear systems using direct data-driven techniques. 
The data-driven invariant sets are used to design a data-driven command governor that selects a command for the closed-loop system to enforce constraints. 
Using basis functions, we solve a semi-definite program to learn a sum-of-squares Lyapunov-like function whose unity level-set is a constraint admissible positive invariant set, which determines the constraint admissible states as well as input commands.
Leveraging Lipschitz properties of the system, we prove that tightening the model-based design ensures robustness of the data-driven invariant set to the inherent plant uncertainty in a data-driven framework. 
To mitigate the curse-of-dimensionality, we repose the semi-definite program into a linear program. 
We validate our approach through two examples: First, we present an illustrative example where we can analytically compute the maximum positive invariant set and compare with the presented data-driven invariant set. Second, we present a practical autonomous driving scenario to demonstrate the utility of the presented method for nonlinear systems. 
\end{abstract}
\end{frontmatter}
\section{Introduction}
Safety-critical cyber-physical systems, such as autonomous vehicles, surgical systems, and industrial robots, must operate within strict constraints to avoid damaging the machine or its environment, including the human operators. These constraints are highly application-oriented. Thus, the existing controller may not be primarily designed for such constraints and be only responsible for general issues like stability and reference tracking. For example, the primary controller in an assisted-driving vehicle may be designed for stability, while driver's maneuvers can cause the car to cross lanes. One solution is creating a constraint-enforcing controller (e.g. \ac{mpc}) to replace the existing controller. This is often undesirable since it discards the careful testing and tuning of the existing controller. Furthermore, control design techniques such as \ac{mpc} typically require a model which may not be available. 
In the assisted-driving example, an uncertain model may lead to impulsive maneuvers that destabilize the system e.g., cause the car to roll, or at least discomfort the passengers \cite{zheng20223dop}.
Alternatively, a {\ac{cg}} \cite{garone2017reference} can be added to the existing closed-loop system as an outer-loop constraint-enforcing controller that adjusts the command given to the existing controller. This is advantageous since redesigning the existing controller for every new scenario is challenging, while {\ac{cg}} is implemented as an add-on to accommodate new scenarios. Moreover, the proposed direct data-driven \acp{cg} can be designed without having the information about the system that was required for designing the existing controller.

Solving \acp{cg} can be intractable when the dynamics are mathematically inaccessible, complex, and uncertain. Some formulations of \acp{cg} (e.g.~\cite{masti2020direct}) require infinite horizon planning to ensure the current command is safe, which can be mitigated by using high-level dynamical properties such as invariance \cite{nakano2023explicit}. In this context, a \acf{ci} set is a subset of the state-space for which there exists a command trajectory such that the state trajectory does not leave that set. The \ac{ci} sets guarantees the \textit{existence} of a command that ensures invariance. However, computing the feasible commands is not trivial. Even if the plant model and a \ac{ci} set are mathematically accessible, computing the command may be intractable e.g., require solving a non-convex optimization problem. This is more challenging in a data-driven approach. For data-sets of the form $\{x_{t+1},x_t,r_t\}$, we only have one command $r_t$ associate with each state $x_t$ at each time $t$, which may or may not promote invariance since the \ac{ci} set is not known in advance.

We simplify the problem of computing a \ac{ci} set to the computation of a collection of \ac{pi} sets for different commands whose union is a \ac{ci} set. Once the \ac{pi} sets are computed, they form an admissible set that reduces the {\ac{cg}} to a one step planner; find a command that keeps the state inside the admissible set. When the model of the system is available, the invariant sets can be computed by standard algorithms as in~\cite{bertsekas1972}. These algorithms can compute maximal output admissible sets for linear systems as in~\cite{gilbert1991linear}, that promotes constraint enforcement. This form of {\ac{cg}} is popular for linear systems (e.g.~\cite{wang2022robust, burlion2022reference, mulagaleti2021}) as well as nonlinear systems (e.g.,~\cite{gilbert2002nonlinear}). However, when the model is not available, these algorithms require indirect data-driven approach wherein a model is first identified from data and then used for model-based controller synthesis. 
Moreover, computing the maximal invariant sets may be overly ambitious. 
{Despite the recent advances for linear systems \cite{ossareh2023data}, to the best of our knowledge, a data-driven approach that finds the exact maximal invariant sets for general nonlinear systems does not exist}. We will settle for a sub-maximal invariant set, provided we can guarantee its invariance despite the uncertainty due to data-driven framework. This is challenging since approximations of an invariant set are not approximately invariant, e.g.~\cite{korda2020computing}. Thus, an approximate invariant set may allow constraint violates to occur. 

This paper present an alternative \textit{direct} data-driven approach wherein the invariant set is learned directly from data. 
Learning-based techniques show promise in constructing Lyapunov-like and barrier certificates from data that support invariance. Learning neural Lyapunov functions can either quantify invariance for autonomous systems (e.g.~\cite{abate2020formal, deka2023supervised}) or design stabilizing controllers (e.g.~\cite{chang2019neural}). The learning process usually includes solving a \ac{sdp}, which is often dealt with \ac{sos} programming for polynomial systems with polynomial constraints~\cite{ribeiro2022nonlinear, cotorruelo2021}. Our approach is inspired by \ac{sos} programming but is not limited to polynomial systems. A similar approach is introduced in~\cite{richards2018lyapunov} though they do not discuss how to adjust the command, but to compute positive invariance for a certain controller. Another similar approach in~\cite{osorio2022novel} includes the control design but requires a linear model that represents the nonlinear system at steady state. Furthermore, there exist data-driven {\ac{cg}} methods (e.g.~\cite{masti2020direct}) that do not use invariance principle and try to directly solve a finite-time receding horizon optimization problem using machine learning tools, for which computational cost and tractability as well as formal safety guarantees need to be investigated.

\subsubsection*{Notations and definitions}
The Frobenius inner-product of square matrices $A,B$ is denoted by $\langle A,B\rangle=\text{Tr}(A^\trans B)$, where Tr$(.)$ is the trace operator. Matrix inequality $A\succ B$ means $A-B\succ0$ is positive definite. The successor of the state $x_t$ is denoted by $x_t^+=x_{t+1}$. {The one-step predecessor of a set $\C$ is defined $\textnormal{Pre}(\C)=\{x_t:x_t^+\in\C\}$}. The argument(s) of the functions and time index $t$ are sometimes dropped to simplify the notation.
The $\circ$ denotes composition. 

For an autonomous system $x^+ =f(x)$, a set $\O$ is \ac{pi} if for any $ x\in\O$ we have $f(x) \in \O$.
For a controlled system $x^+ =f(x,r)$, a set $\C$ is \ac{ci} if for any $ x\in\C$ there exists a command $r\in\R$ such that $f(x, r) \in \C$. 
The max-\ac{pi} set $\Oinf$ (max-\ac{ci} set $\Cinf$) is the \ac{pi} (\ac{ci}) set which contains all other \ac{pi} (\ac{ci}) sets $\O \subseteq \Oinf$ ($\C \subseteq \Cinf$) that are subsets of the constraint set $\O \subseteq \X$ ($\C \subseteq \X$).

\section{Command governor synthesis problem}
\label{sec:problem}
Our objective is to use data to perform constrained command tracking for an unmodeled system. Our knowledge about the unmodeled system is limited to a data-set comprised of state trajectories corresponding to several {constant commands} $\bar r\in\reals$. For the purpose of proving invariance, we assume that the data comes from an unmodeled discrete-time dynamical system 
\begin{equation}
   \label{eq:system}
    x^+ =f(x, r), \quad x\in\X,~r\in\R,\\
\end{equation}
where $f:\reals^n\times\reals\mapsto\reals^n$ is Lipschitz continuous, $r\in\R$ is the command in the range $\R\subseteq\reals$, and $x\in \X\subseteq \reals^n$ is the state that is aimed to stay in the state constraint set
\begin{equation}
\label{eq:constraint}
    \X:=\{x:~g(x)\le1\},
\end{equation}
which is expressed as a sub-level set of some Lipschitz continuous function $g:\reals^n\mapsto\reals$. {We note that measurement noise and external disturbances are not considered here but will be subject of future work.} We use data
\begin{equation}
\label{eq:data}
    \D=\big\{\{ \{x_t^{j,i}\}_{t=0}^T\}_{j=1}^{N_T},~ \bar r^i\big\}_{i=1}^{N_{\bar r}},
\end{equation}
{where $x_t^{j,i}$ denotes the state trajectory, generated from the $j$-th initial condition $x_0^{j,i}$ using the $i$-{th} {constant command} $r=\bar r^i$. In total, the dataset comprises $N_{\bar r}$ {constant commands} $\bar r^i$, each associated with $N_T$ state trajectories generated from different initial conditions. For notational simplicity, both the length $T+1$ and the number $N_{\bar r}$ of trajectories are assumed constant for all commands.} We pose the following assumptions about the unmodeled system~\eqref{eq:system}.
\begin{assumption}
\label{assumption:system}
(System assumptions)
\begin{enumerate}[(a)]
    \item The state $x$ is available, but the dynamics $f$ are unmodeled.\label{assumption:dynamics}
    \item The system~\eqref{eq:system} is asymptotically \ac{iss}.\label{assumption:iss}
    \item The dynamics $f$ are Lipschitz continuous with known Lipschitz constant $L_f$. \label{assumption:lipschitz}
\end{enumerate}
\end{assumption}
Assumption~\ref{assumption:system}\eqref{assumption:dynamics} generally holds for dynamical systems where the state $x$ is measurable. Alternatively, we can define the state $x$ to be a window of any \emph{flat} measurement $\{y_k\}_{k=t-n+1}^t$ of the system when the unmodeled dynamics $f$ is \emph{differentially flat}~\cite{diwold2021normal}.
Assumption~\ref{assumption:system}\eqref{assumption:iss} ensures that for a {constant command} $r_t=\bar r$, the state converges to a unique equilibrium $x_t\to x_\infty(\bar r)$ as $t\to\infty$. This holds, for example, when the system is equipped with an inner-controller.
There are several reasons that the closed-loop dynamics~\eqref{eq:system} could be unmodeled despite the existence of the inner-controller:

\begin{itemize}
    \item {The inner-controller was designed using model-free heuristic e.g. a hand-tuned PID controller, where the plant model was not available.}
    \item The plant was manufactured by a third party that is unwilling to share their proprietary models.
    \item The inner-controller is a complicated collection of legacy controllers and switching logic.
    \item A high-fidelity model is available but is too high-dimensional or complex for analytic computation.
    \item {The inner-loop dynamics include factors that are hard to model, e.g. human-in-the-loop or turbulent effects.}
\end{itemize}
Assumption~\ref{assumption:system}\eqref{assumption:lipschitz} will be used to guarantee robustness to plant uncertainty due to the data-driven framework, since we have finite data~\eqref{eq:data} rather than a high-fidelity model~\eqref{eq:system}. {The Lipschitz constant $L_f$ can be computed from data~\cite{chakrabarty2020safe}}. We note that all items of Assumption~\ref{assumption:system} are consistent with recent literature~\cite{osorio2022novel}.

\subsection{Command governors}
\tikzstyle{block} = [draw, rectangle, minimum height=3em, minimum width=4em]
\tikzstyle{block2} = [draw, rectangle, minimum height=3em, minimum width=4em]
\tikzstyle{input} = [coordinate]
\tikzstyle{output} = [coordinate]
\tikzstyle{feedback} = [coordinate]
\tikzstyle{pinstyle} = [pin edge={to-,thin,black}]
\begin{figure}
    \centering
\begin{tikzpicture}[auto, node distance=1cm,>=latex']
    \node [input, name=input] {};
    \node [block2, right of=input, node distance=1.5cm] (rg) {{\ac{cg}}};
    \node [block, right of=rg, node distance=2.7cm] (controller) {Inner-Controller};
    \node [block, right of=controller, node distance=2.5cm] (system) {Plant};
    \node [output, right of=system] (output1) {};
    \node [output, right of=system,node distance=1.5cm] (output2) {};
    \node [feedback, below of=controller, node distance=1cm](feedback){};
    \node [feedback, below of=feedback, node distance=1cm] (memory) {};

    \draw [->] (input) -- node {$r_d$} (rg);
    \draw [->] (rg) -- node[name=r, near start] {$r$} (controller);
    \draw [->] (controller) -- node[name=u]{$u$} (system);

    \draw [-] (system) -- node [name=y1] {}(output1);
    \draw [->] (output1) -- node [name=y] {$x$}(output2);
    \draw [-] (y1) |- node [near end, above] {Inner-loop}(feedback);
    \draw [->] (feedback) -| node {}(controller);
    \draw [-] (y) |- (memory);
    \draw [->] (memory) -| node[near start, above] {Outer-loop} (rg);
    \node[draw, dashed, inner xsep=3mm,inner ysep=3mm,fit=(controller)(system)(feedback),label={$x^+=f(x,r)$}]{};
\end{tikzpicture}
    \caption{\Acf{cg} design for a system equipped with an inner-controller.}
    \label{fig:rg}
\end{figure}
A {\acf{cg}} is an outer-loop controller, as shown in Fig.~\ref{fig:rg}, that minimally adjusts the command $r$ from the desired command $r_d$ to ensure the constraints $\X$ of the system~\eqref{eq:system} are not violated $x\in\X$. Often this is posed as a dynamic constrained optimization problem with known dynamics $f$ for prediction. Alternatively, a {\ac{cg}} can be reduced to a static optimization problem, if an admissible set of the system~\eqref{eq:system} is available. Then, {\ac{cg}} can be posed as
\begin{subequations}
\label{eq:cg}
\begin{align}
    r\in\arg\min_{\tilde{r}}\> &|\tilde{r}-r_d| \label{eq:cg-cost}\\
    \text{s.t.}~ &{(x, \tilde{r})} \in \Lambda,
\end{align}
where $\Lambda$ is an admissible set
\begin{equation}
\label{eq:cg-Lambda}
    \Lambda\!=\!\{(x, r):~r\!\in\!\R, ~f(x,r)\in\C\},
\end{equation}
\end{subequations}
which is derived from a \ac{ci} set $\C\subseteq\X$ of the system~\eqref{eq:system}. When {\ac{cg}}~\eqref{eq:cg} is recursively feasible $x_t\in\C\subseteq\X,\forall t$, constraint enforcement is guaranteed. The following lemma shows that the existence of a \ac{ci} set is essential for recursive feasibility of {\ac{cg}}~\eqref{eq:cg}. 
\begin{lemma}
\label{lemma:rg-feasible} {\ac{cg}}~\eqref{eq:cg} is recursively feasible for all the states $x\in\textnormal{Pre}(\C)$ if and only if $\C$ is \ac{ci}.
\end{lemma}
\begin{pf}
    Since $x\in\textnormal{Pre}(\C)$, by definition there exist a command that takes the state successor $f(x,r)$ to $\C\subseteq\X$. If $\C$ is \ac{ci}, then there exist a command at any time that keeps the state in $\C\subseteq\X$. If $\C$ is not \ac{ci}, then there exists a point $x\in\C$ that state trajectory starting from $x$ will eventually leave $\C$. $\qed$
\end{pf}
Lemma~\ref{lemma:rg-feasible} suggests that the sufficient plant knowledge required for synthesizing a recursively feasible {\ac{cg}}~\eqref{eq:cg} is only an admissible set $\Lambda$~\eqref{eq:cg-Lambda}, which includes a \ac{ci} set $\C\subseteq\X$ and the range of the command $r$ for every point in $\C$ that renders $\C$ being \ac{ci}. 
We note that {\ac{cg}}~\eqref{eq:cg} {is the least restrictive} when the max-\ac{ci} set $\Cinf$ is used, which produces the maximal admissible set $\Lambda_\infty$ of the system~\eqref{eq:system} for~\eqref{eq:cg-Lambda}. In the next section we will show that our approximation $\Lambda\subseteq \Lambda_\infty$ is an admissible set since it will be derived from a \ac{ci} set $\C\subseteq \Cinf$. Finally, finding the feasible command $r$ that satisfy $f(x,r)\in\C$ may not be tractable even if the model $f$ and the \ac{ci} set $\C$ are known, e.g.~\eqref{eq:cg} may be a non-convex optimization.
In contrast, our approach computes both the \ac{ci} set $\C$ and the feasible commands $r$.

\section{Data-driven command governor}
\label{sec:data-driven}
In this section, we first describe a model-based method to compute a \ac{ci} set. Then, we extend the results to a data-driven framework. 

\subsection{{PI approximations of CI sets}}
The definition of a \ac{ci} set is based on the \textit{existence} of a command $r$ such that $f(x,r) \in \C$. This is fundamentally challenging for data-driven invariance since our data-set may not contain the appropriate commands. In this section, we overcome this issue by inner-approximating a \ac{ci} set using a union of \ac{pi} sets which are more amenable to data-driven synthesis. 

We use the fact that once the {constant command} $r_t= \bar r$ is determined for all time $t$, the system~\eqref{eq:system} becomes an autonomous system with a particular dynamics $f(x; \bar r)$ parameterized by the command $\bar r$. Now, we can define \ac{pi} sets for the autonomous dynamics $f(x; \bar r)$, and use these \ac{pi} sets to create a \ac{ci} set for the system~\eqref{eq:system}. This is true since every \ac{pi} set $\O({\bar r}){\subseteq}\X$ of the autonomous dynamics $f(x; \bar r)$ parameterized by the command $\bar r\in \R$ is a subset of the max-\ac{ci} set $\Cinf$ of the system~\eqref{eq:system}. The following proposition shows that the union 
\begin{equation}
\label{eq:union}
    \C=\bigcup_{\bar r\in \bar \R}\O({\bar r})\subseteq \Cinf \subseteq \X
\end{equation}
of several \ac{pi} sets $\O({\bar r})$ over any subset $\bar \R \subseteq \R$ is a \ac{ci} set, {which is sufficient to create the admissible set \eqref{eq:cg-Lambda}}.
\begin{proposition}
\label{proposition:union}
    Let $\O(\bar r)$ be \ac{pi} for the autonomous dynamics $f(x; \bar r)$ where $\bar r\in \bar \R$. Then,~\eqref{eq:union} is \ac{ci} for the system~\eqref{eq:system}.
\end{proposition}
\begin{pf}
    For the state $x\in\C$, we have $x\in \O(\bar r)$ by some $\bar r\in \bar \R$. Therefore, the command $\bar r$ keeps the state successor $x^+\in \O(\bar r)$, thus renders~\eqref{eq:union} \ac{ci}. $\qed$ 
\end{pf}
The advantage of~\eqref{eq:union} is that it simplifies the problem of computing a \ac{ci} set $\C$ to the computation of \ac{pi} sets $\O({\bar r})$. This is especially helpful in data-driven approach since the \ac{ci} set~\eqref{eq:union} includes all admissible commands supported by data, where the corresponding admissible state regions are described by the \ac{pi} sets $\O({\bar r})$. 

As a result of Proposition~\ref{proposition:union}, {\ac{cg}}~\eqref{eq:cg} can be revised by
\begin{subequations}
\label{eq:cg-aprox}
\begin{align}
    i\in\arg\min_{i}\> &|\bar r^i-r_d|\\
    \text{s.t.}~ &\begin{bmatrix} x\\\bar r^i \end{bmatrix} \in \bar \Lambda,
\end{align}
where the admissible set
\begin{equation}
    \label{eq:cg-aprox-Lambda}
    \bar \Lambda=\{(x, \bar r): \bar r\in\bar \R~\text{s.t.}~ x\in\O({\bar r})\}
\end{equation}
\end{subequations}
enables the synthesis of {\ac{cg}} after the \ac{pi} sets $\O({\bar r})$ are computed for every {constant command} ${\bar r}\in \bar \R$. The new representation~\eqref{eq:cg-aprox} of {\ac{cg}} is now an \ac{ip}, which selects between the commands ${\bar r}$ that enforces constraints $\X$. Solving {\ac{cg}}~\eqref{eq:cg-aprox} finds the nearest command $\bar r \in \bar \R$ such that $x_t\in \O(\bar r)$ the current state $x_t$ lies in the corresponding \ac{pi} set $\O(\bar r)$. {The computational complexity of {\ac{cg}}~\eqref{eq:cg-aprox} can be mitigated by leveraging scalar optimization techniques e.g. decision tree, branch and bound, and bisection.}
The following corollary of Lemma~\ref{lemma:rg-feasible} shows {\ac{cg}}~\eqref{eq:cg-aprox} is recursively feasible.
\begin{corollary}
    \label{corollary:feasibilityO}
    The {\ac{cg}}~\eqref{eq:cg-aprox} is recursively feasible {for all the states $x\in\textnormal{Pre}(\C)$} if $\O({\bar r})$ is \ac{pi} for every command ${\bar r}\in \bar \R$, where $\C$ was defined in \eqref{eq:union}.
\end{corollary}
\begin{pf}
The proof follows the proof of Lemma~\ref{lemma:rg-feasible}, where $\C\subseteq\X$ is \ac{ci} by Proposition~\ref{lemma:rg-feasible}, and $x\in\C$ is guaranteed for all time since $\O(\bar r)$ is \ac{pi}. $\qed$
\end{pf}

\subsection{{Functional synthesis of PI sets}}
For the \ac{pi} sets $\O(\bar{r})$, we use the sub-level representation 
\begin{equation}
\label{eq:pi-set}
    \O({\bar r})=\{x:~V(x;\bar r) \le1\}
\end{equation}
of some function $V$ to transform the problem of synthesizing the set $\O({\bar r})\subseteq\X$ into the problem of synthesizing the function $V$. This allows us to leverage machine learning techniques such as neural Lyapunov-like functions~\cite{mathiesen2022safety,chang2019neural} to create the \ac{pi} set $\O(\bar r)$.
To achieve this, a \textit{function approximation architecture} is often fixed and its parameters are tuned. We consider a combination of {pre-selected} basis functions 
\begin{equation*}
    \phi(x)=[\phi_1(x),\cdots, \phi_{N_\phi}(x)]^\trans\in\reals^{N_\phi}, 
\end{equation*} 
where the scalar functions $\phi_i(x),~ i=1,\cdots,N_\phi,$ are Lipschitz continuous. 
{We use the basis functions $\phi$ to parameterize the \ac{pi} sets $\O({\bar r})$, as well as an inner-approximation $\hat{\X}\subseteq\X$ of the constraint set $\X$.}
Indeed, the quality of the approximations depend on the choice of the basis functions $\phi(x)$. For instance, {~\cite{beatson2011better} evaluates the fitting quality of radial basis functions by considering their completeness and generalization properties}. 
{To inner-approximate the constraints $\mathcal{X}$, we consider a linear combination $c^\trans \phi$ of the basis functions
}
\begin{equation*}
\label{eq:xhat}
   \hat \X=\{x:~c^\trans \phi(x)\le1\}\subseteq\X
\end{equation*}
where $c\in \reals^{N_\phi}$ is the solution to the \ac{lp}
\begin{subequations}
\label{eq:coef-con}
    \begin{align}
            c=\arg\min_{\tilde c}\> ~~~&\tilde c^\trans \int_\X \phi(x) ~\mathrm{d}\mu(x), \label{eq:coef-con-cost}\\
    \label{eq:coef-con-sub} \text{s.t.  } &  \tilde c^\trans \phi(x) \ge g(x),~~\forall x, 
    \end{align}
\end{subequations}
and $g(x)$ characterizes the constraint set $\X$ from~\eqref{eq:constraint}.
{The cost~\eqref{eq:coef-con-cost} is the Lebesgue integral with respect to some measure $\mu$ over $\X$, which maximizes the volume of the inner-approximation $\hat{\X}\subseteq\X$ and can be pre-computed independently of the decision variable $\tilde{c}$. We note that the feasibility of \ac{lp}~\eqref{eq:coef-con} depends on the generalization properties of the basis functions $\phi$~\cite{beatson2011better}.}
{However, we can avoid solving the \ac{lp}~\eqref{eq:coef-con} by adding $g(x)$ as an element of $\phi(x)$, which gives the exact representation $\hat \X=\X$. 
For instance, $c=[1,0,\cdots,0]^\trans$ is the trivial solution to~\eqref{eq:coef-con} when $\phi_1(x)=g(x)$.} 
{We will later use $c$ to ensure constraint admissibility of our \ac{pi} sets.}

For the sets $\O({\bar r})$ \eqref{eq:pi-set} to be constraint admissible and \ac{pi}, the function $V$ must satisfy the following conditions
\begin{subequations}
    \label{eq:Oinf-condition}
    \begin{align}
         V(x~;\bar r)> 1~~~~&\forall x\not\in\hat{\X} ,\label{eq:Oinf-admissible}\\
         V(f(x;\bar r)~; \bar r)\le 1 ~~~~&\forall x\in\O(\bar r), \label{eq:Oinf-invariant}
    \end{align}
\end{subequations}
{where \eqref{eq:Oinf-admissible} ensures $\O\subseteq\hat{\X}$ by the S-procedure, and~\eqref{eq:Oinf-invariant} ensures $\O$ is \ac{pi}}.
{Although condition~\eqref{eq:Oinf-condition} is effective for evaluating the invariance of a candidate set $\O({\bar r})$, it is not amenable for synthesis since~\eqref{eq:Oinf-invariant} depends on the unknown \ac{pi} set $\O({\bar r})$.}
Thus, similar to Lyapunov-like and barrier functions approaches, we relax the condition~\eqref{eq:Oinf-condition},
\begin{subequations}
    \label{eq:Oinf-lyap}
    \begin{align}
        V(x~;\bar r)&> 1 &\forall x\!\not\in\!\hat{\X},\label{eq:lyap-admissible}\\
        V(f(x;\bar r)~;\bar r)\!&\le \!(1\!-\!\gamma) ~V\!(x~;\bar r)+\gamma \!-\! \beta &\forall x\!\in\!\hat{\X}, \label{eq:lyap-invariant}
    \end{align}
\end{subequations}
where {the Lyapunov-like condition~\eqref{eq:lyap-invariant} is enforced on $\hat{\X}\subseteq\X$ which implies~\eqref{eq:Oinf-invariant} on the unknown \ac{pi} set $\O(\bar r)$}.
{Although the model-based condition~\eqref{eq:Oinf-condition} did not require further assumptions on $V$, we will impose Lipschitz continuity on $V$ for the relaxed conditions~\eqref{eq:Oinf-lyap}.}
{The parameters $\gamma \in [0,1]$ and $\beta \in [0,\gamma]$ are introduced to relax the Lyapunov decrease condition~\eqref{eq:lyap-invariant}. The parameter $\gamma$ creates a convex combination $(1\!-\!\gamma)V+\gamma$ of $V$ and $1$ in~\eqref{eq:lyap-invariant}. This dictates that $V$ must be decreasing $V^+=V(f(x;\bar r)~;\bar r)\le V$ outside $\O$ i.e. $V^+ \leq (1-\gamma)V + \gamma \leq V$ when $V > 1$. But, it allows $V$ to be increasing inside $\O$ while $V^+ \leq (1-\gamma)V + \gamma \leq 1 $. The drift parameter $\beta$ enforces that $V$ is decreasing in the region $\{x:~V\!\ge\!\frac{\gamma-\beta}{\gamma}\}$ that includes an annulus around the boundary of $\O$. Furthermore, we will use $\beta$ to provide robustness to the uncertainty due to data-driven framework.}
We parameterize the Lyapunov-like function as 
{\begin{equation}
\label{eq:lyapunov-function}
    V(x~;\bar r)\!=\! \langle P, \varphi(x;\bar r)\varphi(x;\bar r)^\trans\rangle +2 c^\trans \phi_\infty\!-(c^\trans \phi_\infty)^2\!,
\end{equation}
}where $x_\infty \in \hat\X$ is the equilibrium of \eqref{eq:system} for constant command $\bar r$, $\phi_\infty=\phi(x_\infty(\bar r))$, and $\varphi(x~;\bar r)=\phi(x)-\phi_\infty$. {The positive definite matrix $P\in\reals^{N_\phi\times N_\phi} \succ 0$ is the decision variable to be tuned}, and $c$ is the solution to \eqref{eq:coef-con}. The function $V$ is a quadratic combination of the basis functions $\phi$ due to the first term $\langle P, \varphi\varphi^\trans\rangle=\varphi^\trans P \varphi$. 
{The Lyapunov-like function~\eqref{eq:lyapunov-function} is only valid for feasible equilibria $x_\infty \in \hat\X$ where $c^\trans \phi_\infty \leq 1$. Otherwise, we define $\O(\bar r) = \varnothing$ for $x_\infty \not\in\hat\X$.}
\subsection{Model-based computation of PI sets}
Before presenting our data-driven approach, we first present the results for the case where the autonomous dynamics $f(x;\bar r)$, and thus, the corresponding equilibrium $x_\infty(\bar r)$ are known. In this case, {we encode the condition \eqref{eq:Oinf-lyap} as the following \ac{sos} program}
\begin{subequations}
\label{eq:model-sdp}
\begin{align}
   P \!\in \! \arg\min_{\tilde P} & \langle \tilde P, \Psi \rangle \label{eq:model-cost}\\
    \text{s.t. } & \langle \tilde P, \psi \rangle \le \gamma_0,~~~\forall x\in\hat{\X} \label{eq:model-decrease}\\
    & cc^\trans \preceq \tilde P, \label{eq:model-admissible}
    \end{align}
\end{subequations}
where $\psi=(\varphi \circ f)(\varphi \circ f)^\trans-(1-\gamma) \varphi\varphi^\trans$, the $\gamma_0=\gamma(1-c^\trans \phi_\infty)^2-\beta$ is independent of $x$, and
\begin{equation}
    \label{eq:psi}
    \Psi=\int_{\hat\X}\! \varphi \varphi^\trans~\mathrm{d} \mu.
\end{equation}
The cost \eqref{eq:model-cost} minimizes a Lebesgue integral of the Lyapunov-like function~\eqref{eq:lyapunov-function} with respect to some measure $\mu$ over $\hat{\X}$, since the last two terms of $V$ from~\eqref{eq:lyapunov-function} are constant. This rewards enlarging the level-sets of $V$, and hence $\O(\bar r)$. 
The following proposition shows the solution to the model-based \ac{sdp}~\eqref{eq:model-sdp} renders $\O(\bar r)$ constraint admissible \ac{pi}.
\begin{proposition}
\label{proposition:model-based}
    Let the dynamics $f$, and equilibrium $x_\infty\in\hat{\X}$ be known. Let $c$ and $P$ be feasible solutions to~\eqref{eq:coef-con} and~\eqref{eq:model-sdp}, respectively. Then, $\O({\bar r})$ from~\eqref{eq:pi-set} is (i) constraint admissible $\O({\bar r})\subseteq \hat{\X}$, and (ii) \ac{pi} for the autonomous dynamics $f(x;\bar r)$ of the system~\eqref{eq:system} with command $r_t=\bar r$.
\end{proposition}
\begin{pf}
First, we show $\O(\bar r)$ is (i) constraint admissible $\O({\bar r}) \subseteq \X$. 
{The lower bound~\eqref{eq:model-admissible} on $P$ enforces the scalar inequality $V\ge(c^\trans\varphi)^2+2c^\trans\phi_\infty-(c^\trans\phi_\infty)^2$ on \eqref{eq:lyapunov-function}. Following the S-procedure, when $x\in\O({\bar r})$, we have $V\le1$ which yields $(c^\trans\varphi)^2+2c^\trans\phi_\infty-(c^\trans\phi_\infty)^2\le 1$ because~\eqref{eq:model-admissible} is enforced. Rearranging the terms gives $(c^\trans \varphi)^2\le(1-c^\trans\phi_\infty)^2$, which implies $c^\trans \varphi\le1-c^\trans\phi_\infty$ where $c^\trans\phi_\infty\le1$ by the assumption $x_\infty\in\hat{\X}$. By the definition $\varphi=\phi-\phi_\infty$, the last inequality yields $c^\trans\phi\le1$ which indicates $x\in\hat{\X}$. Thus $\O({\bar r})\subseteq \hat{\X}\subseteq\X$.} 

Next, we will show $\O({\bar r})$ is \ac{pi}. Condition \eqref{eq:model-decrease} is equivalent to \eqref{eq:lyap-invariant}, where we used the linearity of inner-products for the left hand side and completed the square of the constants to form $\gamma_0$ on the right hand side of \eqref{eq:model-decrease}. Thus, for every $x\in\O({\bar r})$ where $V\le1$ we have $V\circ f\le1$ by \eqref{eq:lyap-invariant}, which implies the invariance condition \eqref{eq:Oinf-invariant}. $\qed$
\end{pf}
The choice of the basis $\phi$ plays the key role in enforcing~\eqref{eq:model-decrease}. 
Classic \ac{sos} programming uses polynomial basis functions $\phi$ where the dynamics $f$ belong to the space of polynomials~\cite{cotorruelo2021}. Kernel methods in \ac{sos} programming use functional polynomials to capture nonlinearities, where the dynamics become polynomials after a nonlinear transformation~\cite{anderson2015advances}. 
In our numerical results, we use machine-learning basis functions to create a \ac{sos} Lyapunov-like shallow neural network \cite{richards2018lyapunov}.

\subsection{{Data-Driven computation of PI sets}}

In this section, we adapt the model-based \ac{pi} set synthesis problem~\eqref{eq:model-sdp} to a data-driven paradigm. Algorithm~\ref{alg:mci} summarizes our data-driven method for calculating the \ac{ci} set~\eqref{eq:union}. It uses the data-set~\eqref{eq:data} to compute the \ac{pi} sets $\O(\bar r)$~\eqref{eq:pi-set} for every command $\bar r\in\bar \R$. The main operation in Algorithm~\ref{alg:mci} is computing $P$ from the data~\eqref{eq:data}. This requires transforming the model-based method~\eqref{eq:model-sdp} into a data-driven method, {which presents two challenges}. First, the equilibrium point $x_\infty(\bar r)$ is not necessarily available. Thus, we use an ensemble average to estimate
\begin{equation}
    \label{eq:xinf}
    x_\infty(\bar r^i) \approx \frac{1}{N_T}\sum_{j=1}^{N_T}{x_T^{j,i}}
\end{equation}
for every $\bar r\in \{\bar r^i\}_{i=1}^{N_{\bar r}}$, where $x_T^{j,i}$ is the last sample of the $j$th trajectory from the data $\eqref{eq:data}$ that is generated by the command $r_t=\bar r^i$.
Assumption~\ref{assumption:system}\eqref{assumption:iss} ensures that every constant command $\bar r$ takes the system to the unique equilibrium $x_T \to x_\infty(\bar r)$, asymptotically as $T\to \infty$. We will prove that our \ac{pi} sets are robust to the approximation~\eqref{eq:xinf} with finite $T$. Second, since we only have finite data, our plant is uncertain. The condition~\eqref{eq:model-decrease} must hold for every state $x \in \hat{\X}$, while in a data-driven paradigm only a finite number $N_s$ of sample pairs $\{(x_k^+, x_k)\}_{k=1}^{N_s}$ are available. Thus, to insure the \ac{pi} set $\O(\bar r)$~\eqref{eq:pi-set} is robust to plant uncertainty due to the data-driven framework, {we tighten~\eqref{eq:model-decrease} by $\epsilon_k>0$, where $\epsilon_k$ satisfies}
\begin{equation}
    \label{eq:epsilon}
    \epsilon_k \ge 2L_{\phi} \delta(L_f\|\varphi_k^+\|+\|\varphi_k\|)+ (L_{\phi} L_f\delta)^2,
\end{equation}
and $L_{\phi}$ and $L_f$ are the Lipschitz constants of the basis functions $\phi$ and dynamics $f$, respectively. {The data density $\delta$ is the Hausdorff distance between the constraint set $\X$ and data-set $\D$, which can be computed using e.g.~\cite{har2015net}.}
Similarly, the Lipschitz constant $L_f$ can be computed from data using e.g.~\cite{chakrabarty2020safe}, and $L_{\phi}$ is known for given basis functions $\phi$. The function $\varphi$ from~\eqref{eq:lyapunov-function} is used to ``lift'' the sample pairs $\{(x_k^+, x_k)\}_{k=1}^{N_s}$,
\begin{equation}
\label{eq:phi-lifted}
   \varphi_k=\varphi(x_k;\bar r), ~~~ \varphi_k^+=\varphi(x_k^+;\bar r).
\end{equation}
Our data-driven approach finds the matrix $P$ defining the \ac{pi} sets $\O(\bar r)$~\eqref{eq:pi-set} {for each $\bar r$} by the following \ac{sdp}
\begin{subequations}
\label{eq:data-sdp}
\begin{align}
   P \in \arg\min_{\tilde P} ~ & \langle \tilde P, \Psi \rangle \label{eq:data-sdp-cost}\\
    \text{s.t.  } &\langle \tilde P,\psi_k \rangle \le \gamma_0-\epsilon_k\lambda,~ k\!=\!1,\!\cdots\!,\!N_s,\label{eq:data-decrease}\\
    & cc^\trans \preceq \tilde P\preceq \lambda I_{N_\phi}, \label{eq:data-admissible}
    \end{align}
\end{subequations}
where $\psi_k=\varphi_k^+\varphi_k^{+\trans} -(1-\gamma)\varphi_k\varphi_k^\trans$, and $\epsilon_k$ satisfies~\eqref{eq:epsilon}. We may either treat $\lambda$ as a decision variable or pre-assign it in \ac{sdp}~\eqref{eq:data-sdp}. Furthermore, $\Psi$ in the cost~\eqref{eq:data-sdp-cost} can be analytically calculated by~\eqref{eq:psi} since $\varphi=\phi-\phi_\infty$ is available in the analytical form. {Alternatively, we can replace the analytic integral $\Psi$ in~\eqref{eq:psi} with the data-driven numerical integral}
\begin{equation*}
   \Psi=\sum_{k=1}^{N_s} \varphi_k\varphi_k^\trans \mu(x_k),
\end{equation*}
with some measure $\mu$.
\begin{algorithm}[t]
\caption{Data-driven \ac{ci} set computation}
\label{alg:mci}
\begin{algorithmic}[1]
\STATE \textbf{input:} Data set $\D$ from~\eqref{eq:data}.
\FOR {$i=1,\cdots,N_{\bar r}$}
    \STATE approximate the equilibrium $x_\infty(\bar r^i)$ by~\eqref{eq:xinf}
    \FOR{$k=1,\cdots,N_s$, where $x_k\in\D^i$}
        \STATE lift \begin{align*}
            \varphi_k&=\phi(x_k)-\phi(x_\infty(\bar r^i)) \\
            \varphi_k^+&=\phi(x_k^+)-\phi(x_\infty(\bar r^i))
        \end{align*}
    \ENDFOR
    \STATE solve \ac{sdp}~\eqref{eq:data-sdp} (or \ac{lp}~\eqref{eq:lp-alpha} then~\eqref{eq:lp-P}) for $P$,
    \STATE store $\D_\O^i:=\{x_\infty(\bar r^i), P^i=P\}$,
\ENDFOR
\STATE \textbf{output:} $\{\D_\O^i\}_{i=1}^{N_{\bar r}}$
\end{algorithmic}
\end{algorithm}
The following theorem shows that the solution $P$ to \ac{sdp}~\eqref{eq:data-sdp}, if feasible, produces a \ac{pi} set $\O({\bar r})\subseteq \hat{\X}$~\eqref{eq:pi-set} for the autonomous dynamics $f(x;\bar r)$. We will later demonstrate the feasibility of \ac{sdp}~\eqref{eq:data-sdp}.

\begin{theorem}
\label{theorem:mpi}
Let Assumption~\ref{assumption:system} hold and $x_\infty\in\hat{\X}$ be known. Let $c$ and $P$ be the solutions to~\eqref{eq:coef-con} and~\eqref{eq:data-sdp}, respectively. Let the data samples have a density of $\|x-x_k\| \le \delta$. Then, $\O({\bar r})$ from~\eqref{eq:pi-set} is (i) constraint admissible $\O({\bar r})\subseteq \hat{\X}$, and (ii) \ac{pi} for the autonomous dynamics $f(x;\bar r)$ of the system~\eqref{eq:system} with command $r_t=\bar r$.
\end{theorem}
\begin{pf}
{The proof of (i) follows the proof of Proposition~\ref{proposition:model-based} since $P\succeq cc^\trans$ is carried over from \eqref{eq:model-admissible} to \eqref{eq:data-admissible}. 
To prove (ii) (i.e. $\O({\bar r})$ is \ac{pi}), we need to show \eqref{eq:data-decrease} implies \eqref{eq:model-decrease}. For that, we must bound the uncertainty of the data-driven framework in the first term of the Lyapunov-like function~\eqref{eq:lyapunov-function}, which is purely quadratic $\langle P, \varphi\varphi^\trans\rangle=\|\varphi\|_P^2$ with respect to $\varphi$, where $\|\varphi\|_P=\varphi^\trans P\varphi$ is the weighted 2-norm. This term can be upper-bounded at every state successor $x^+$ with respect to an adjacent sampled state successor $x_k^+$,
\begin{align*}
    \|\varphi^{+}\|_P^2 \!\!&=\|\varphi_k^+ +(\varphi^{+}-\varphi_k^+)\|_P^2\\
    &= \|\varphi_k^+\|_P^2 + 2\varphi_k^{+\trans} P (\varphi^{+}-\varphi_k^+) + \|\varphi^{+}-\varphi_k^+\|_P^2\\
    &\le \|\varphi_k^+\|_P^2 + 2\|\varphi_k^+\|_P \|\varphi^{+}\!-\varphi_k^+\|_P + \|\varphi^{+}\!-\varphi_k^+\|_P^2,
\end{align*}
where $\varphi^+=\varphi(x^+;\bar r)$. 
The last step follows from Cauchy-Schwarz inequality. The term $\|\varphi^{+}\!-\varphi_k^+\|_P$ is bounded by $\|\varphi^{+}\!-\varphi_k^+\|_P\le \lambda \|\varphi^{+}-\varphi_k^+\|\le \lambda L_{\phi}\|f(x;\bar r)-f(x_k;\bar r)\|\le \lambda L_{\phi}L_f\|x-x_k\|\le  \lambda L_{\phi}L_f \delta$, where $\lambda$ bounds the largest eigenvalue of $P$ according to~\eqref{eq:data-admissible}, $L_{\phi}$ and $L_f$ are the Lipschitz constants of the basis functions $\phi$ and dynamics $f$, respectively, and $\delta$ is the data density $\|x-x_k\| \le \delta$. Applying this bound gives the upper-bound
\begin{align}
    \|\varphi^{+}\|_P^2 \le \|\varphi_k^+\|_P^2 + 2 \lambda L_{\phi} L_f\delta\|\varphi_k^+\| + \lambda (L_{\phi} L_f\delta)^2.
        \label{eq:upper}
\end{align}
Similarly, a lower-bound on $\|\varphi\|_P^2$ at every state $x$ with respect to an adjacent sampled state $x_k$ can be derived
\begin{align*}
    \|\varphi\|_P^2 &=\|\varphi_k^+\|_P^2 + 2\varphi_k^{+\trans} P (\varphi^{+}-\varphi_k^+) + \|\varphi^{+}-\varphi_k^+\|_P^2\\
    &\ge \|\varphi_k^+\|_P^2 + 2\varphi_k^{+\trans} P (\varphi^{+}-\varphi_k^+) \\
    &\ge \|\varphi_k\|_P^2 - 2\|\varphi-\varphi_k\|_P\|\varphi_k\|_P,
\end{align*}
where the second line follows from $\|\varphi^{+}-\varphi_k^+\|_P^2 \geq 0$ and the final line follows from the Cauchy-Schwarz inequality.
Again, we bound $\|\varphi-\varphi_k\|_P\le \lambda \|\varphi-\varphi_k\|\le \lambda L_{\phi} \|x-x_k\|\le \lambda L_{\phi} \delta$ to get the lower bound
\begin{equation}
\label{eq:lower}
    \|\varphi\|_P^2 \ge \|\varphi_k\|_P^2 - 2\lambda L_{\phi} \delta\|\varphi_k\|.
\end{equation}
Scaling~\eqref{eq:lower} by $-(1-\gamma)$ and adding it to~\eqref{eq:upper} gives 
\begin{equation*}
    \langle P,\varphi^+\varphi^+\!^\trans \!-\!(1\!-\gamma) \varphi\varphi\!^\trans\rangle\! \le \!\langle P,\varphi_k^+\varphi_k^{+\trans} -(1-\gamma)\varphi_k\varphi_k^\trans \rangle +\epsilon_k\lambda,
\end{equation*}
where $\langle P, \varphi\varphi^\trans\rangle=\|\varphi\|_P^2$, $\epsilon_k$ satisfies~\eqref{eq:epsilon}, and $\varphi^+=\varphi\circ f$. Since \eqref{eq:data-decrease} bounds the right hand side of the latter by $\gamma_0$, thus \eqref{eq:model-decrease} holds for every state $x\in\hat{\X}$. The rest of the proof follows the proof of Proposition~\ref{proposition:model-based}, where we showed $\O({\bar r})$ is \ac{pi} when \eqref{eq:model-decrease} holds.  $\qed$
}
\end{pf} 

Although Theorem~\ref{theorem:mpi} guarantees that the solution of~\eqref{eq:data-sdp} produces a \ac{pi} set, it is not guaranteed that~\eqref{eq:data-sdp} will have a solution. The feasibility of \ac{sdp}~\eqref{eq:data-sdp} depends on several factors including the existence of the Lyapunov-like function in the first place, the generalization properties of the basis functions $\phi$~\cite{beatson2011better}, the data-driven uncertainty compensation $\epsilon_k$, and the accuracy of the equilibrium approximation~\eqref{eq:xinf}. The following theorem shows that \ac{sdp}~\eqref{eq:data-sdp} is feasible if the data~\eqref{eq:data} is sufficiently dense, the length $T+1$ of the sampled state trajectories is sufficiently long, and the lift functions $\phi$ can describe a Lyapunov-like function of the form~\eqref{eq:lyapunov-function} that renders the model-based \ac{sdp}~\eqref{eq:model-sdp} feasible with nonzero $\beta$.

{
\begin{theorem}
\label{theorem:robust}
    Let Assumption~\ref{assumption:system} hold. Let the data samples have a density of $\|x-x_k\| \le \delta$ for a sufficiently small $\delta$. Let an upper-bound $T\ge T_s$ on the settling-time $T_s$ be known such that the system~\eqref{eq:system} has settled $\|x_t-x_\infty (\bar r)\|\le \epsilon_{\bar r}$, $\forall t \ge T_s$ for a sufficiently small $\epsilon_{\bar r}$. Let the model-based \ac{sdp}~\eqref{eq:model-sdp} be feasible for some $\beta=\beta_M\ge0$, Lipschitz continuous $\phi$, and the true equilibrium $x_\infty(\bar r)\in\hat\X$. Then, the data-driven \ac{sdp}~\eqref{eq:data-sdp} is feasible for some $\beta=\beta_D\ge0$, where $x_\infty(\bar r)$ is approximated by~\eqref{eq:xinf}.
\end{theorem}
\begin{pf}
    Let $e(\bar r)= \frac{1}{N_T}\sum_{j=1}^{N_T}{\mathbf{x}_j^i}_T - x_\infty(\bar r)$ denote the equilibrium approximation error for $\bar r=\bar r^i$. We bound $\|\varphi_k\|_P^2=\|\phi(x_k)-\phi(x_\infty(\bar r)+e(\bar r))\|_P^2\le\|\varphi\|_P^2+\lambda L_\phi\epsilon_{\bar r}^2$ by applying the triangle inequality and the Lipschitz continuity of $\phi$, where $\lambda$ bounds the largest eigenvalue of $P$, the error is bounded $\|e(\bar r)\|\le\epsilon_{\bar r}$ by assumption and, with a slight abuse of notation, $\varphi_k$ is computed by using the approximated equilibrium~\eqref{eq:xinf} while $\varphi$ uses the exact equilibrium $x_\infty(\bar r)$. A similar bound can be computed for $\|\varphi_k^+\|_P^2\le\|\varphi\circ f\|_P^2+\lambda L_\phi\epsilon_{\bar r}^2$. Combining the two bounds yields $\langle P,\psi_k \rangle \le \langle P, \psi \rangle +2\lambda L_\phi\epsilon_{\bar r}^2$, where $\psi=(\varphi\circ f)(\varphi\circ f)^\trans -(1-\gamma) \varphi\varphi^\trans$ and $\psi_k=\varphi_k^+\varphi_k^{+\trans} -(1-\gamma)\varphi_k\varphi_k^\trans$.
    Since \eqref{eq:model-sdp} is assumed to be feasible for some $\beta=\beta_M$, the term $\langle P, \psi \rangle$ is bounded by \eqref{eq:model-decrease}, which implies
    $\langle P,\psi_k \rangle \le \gamma(1-c^\trans \phi_\infty)^2-\beta_M +2\lambda L_\phi\epsilon_{\bar r}^2$. Therefore, \eqref{eq:data-decrease} holds when $
        \lambda (2 L_\phi\epsilon_{\bar r}^2+\epsilon_k)\le\beta_M-\beta_D$.
   This holds when, for instance, $\epsilon_k\le\frac{\beta_M-\beta_D}{2\lambda}$ where $\epsilon_k$ satisfies~\eqref{eq:epsilon} with a small enough $\delta<\!\!<\|\varphi_k\|$ i.e. high data density, and small enough $\epsilon_{\bar r}$ such that $\epsilon_{\bar r}^2\le\frac{\beta_M-\beta_D}{4\lambda L_\phi}$.
    Finally, since $\lambda$ bounds the largest eigenvalue of $P$, \eqref{eq:data-admissible} holds when \eqref{eq:model-admissible} holds. Thus, \eqref{eq:data-sdp} is feasible if \eqref{eq:model-sdp} is feasible. $\qed$
\end{pf}
}
The feasibility of the model-based \ac{sdp}~\eqref{eq:model-sdp} and the shape of the \ac{pi} sets $\O({\bar r})$ depends on the choice of the basis functions $\phi$. {For instance, $\phi=x$ creates ellipsoidal invariant sets. For a stable linear system whose constraints contain a neighborhood of the origin, the existence of $P\succ0$ is guaranteed when $\phi=x$ by properties of Lyapunov stability for linear systems. 
}

\subsection{Computational tractability}
{\bf Offline computations:} Solving \ac{sdp}~\eqref{eq:data-sdp} is computationally expensive, especially for a high-dimensional lifting $\phi$. We resolve this by relaxing \ac{sdp}~\eqref{eq:data-sdp} into a {computationally less expensive} \ac{lp} by using positive definite matrices $W_j\in\reals^{N_\phi\times N\phi}, j=1,\cdots,N_W$, where $0\prec W_j\preceq\frac{1}{M}(\lambda I_{N_\phi}-cc^\trans)$, to approximate the solution to the \ac{sdp}~\eqref{eq:data-sdp} by 
\begin{equation}
    \label{eq:lp-P}
    P=cc^\trans + \sum_{j=1}^{N_W} \alpha_j W_j.
\end{equation}
Then, the \ac{sdp}~\eqref{eq:data-sdp} reduces to the following \ac{lp} that solves for the vector $\tilde\alpha=[\alpha_1,\cdots,\alpha_{N_W}]^\trans$,
\begin{subequations}
    \label{eq:lp-alpha}
\begin{align}
    {\alpha}= \arg\min_{\tilde\alpha}~ &  d^\trans{\tilde\alpha} &\\
    \text{s.t.  } & A\tilde\alpha + b \le 0,&\label{eq:lp-decrease}\\
    & 0\le\alpha_j\le1, & j=1,\cdots,N_W,\label{eq:lp-bounds}
\end{align}
\end{subequations}
where the entries of the matrix $A$ and vectors $b$ and $d$ are 
\begin{align*}
    [A]_{k,j}&=\langle W_j,\psi_k \rangle,\\
    [b]_k&= (c^\trans\varphi_k^+)^2 - (c^\trans \varphi_k)^2 -\gamma +\epsilon_k\lambda,\\
    [d]_j&= \langle W_j, \Psi \rangle,
\end{align*} respectively. We note that the vector $d$ is defined such that the costs of \ac{lp}~\eqref{eq:lp-alpha} and \ac{sdp}~\eqref{eq:data-sdp} are equivalent for the $P$ of the form~\eqref{eq:lp-P}. Moreover, the following proposition shows that $P$ from~\eqref{eq:lp-P} is feasible for \ac{sdp}~\eqref{eq:data-sdp}.
\begin{proposition}
\label{proposition:lp-relax-sdp}
    Let $\alpha$ be the solution of~\eqref{eq:lp-alpha}. Then, the matrix $P$~\eqref{eq:lp-P} is a feasible solution for the \ac{sdp}~\eqref{eq:data-sdp}.
\end{proposition}
\begin{pf}
    Condition~\eqref{eq:lp-decrease} enforces $\sum_{j=1}^{N_W}[A]_{k,j} \alpha_j+[b]_k\le0$ for $k=1,\cdots, N_s$, which is equivalent to~\eqref{eq:data-decrease} with $P$ from~\eqref{eq:lp-P}. Condition~\eqref{eq:lp-bounds} enforces $cc^\trans\preceq P\preceq\lambda I_{N_\phi}$~\eqref{eq:data-admissible} by triangle inequality where $0\prec W_j\preceq\frac{1}{M}(\lambda I_{N_\phi}-cc^\trans)$ in~\eqref{eq:lp-P}. $\qed$ 
\end{pf}
Proposition~\ref{proposition:lp-relax-sdp} allows us to avoid the computational complexity of the \ac{sdp}~\eqref{eq:data-sdp}, which is proportional to the square of the lifting dimension $N_\phi^2$. Consequently, we can use more nonlinear functions $\phi$ to capture the nonlinearities of the system in exchange for a sub-optimal, but feasible, solution to \ac{sdp}~\eqref{eq:data-sdp}. Since Theorem~\ref{theorem:mpi} ensures that feasible solutions of~\eqref{eq:data-sdp} are \ac{pi}, the feasible relaxation~\eqref{eq:lp-alpha} also produces provable \ac{pi} sets.

{{\bf Online computations:} Following the big-$\mathbb{O}$ notation, the computational complexity of {\ac{cg}}~\eqref{eq:cg-aprox} is $\mathbb{O}(N_{\bar r}N_\phi^2)$ to verify the set membership $x\in\O({\bar r})$ for every $\bar r\in\bar\R$. This includes $N_{\bar r}$ times computation of $V$ from~\eqref{eq:lyapunov-function}, where the quadratic term $(\varphi^\trans P \varphi)^2$ has a dominating complexity of $\mathbb{O}(N_\phi^2)$, and $N_{\bar r}$ times checking the inequality from \eqref{eq:pi-set}. However, invariance can mitigate the computational complexity. By the recursive feasibility of {\ac{cg}}~\eqref{eq:cg-aprox} from Corollary~\ref{corollary:feasibilityO}, the
solution $r_{t-1}$ is a feasible solution for $r_t$ since it corresponds to a \ac{pi} set $\O(r_{t-1})$. Consequently, the computational complexity reduces to $\mathbb{O}(\tfrac{|r_{t-1} -r_d|}{\Delta r}N_\phi^2)$ if the commands are uniformly discretized by ${\Delta r}$. This is useful when the command changes gradually.}

\section{Numerical examples}
\label{sec:examples}

\subsection{Illustrative example}
To illustrate our data-driven {\ac{cg}}, we consider the following \ac{lti} system,
\begin{equation}
    \label{eq:linear} 
    \dot x =\begin{bmatrix} 0 & 1 \\ -\omega^2 & -2\zeta \omega \end{bmatrix}x + \begin{bmatrix} 0\\ \omega^2 \end{bmatrix} r, ~~x\in\X, ~r\in \R,\\
\end{equation}
where $\X=\{x:[1,0]x\in[-1,1]\}$ is the state constraint, $\R=[-1.2,~1.2]$ is the command range, $\omega=5$ is the natural frequency, and $\zeta=0.1$ is the damping ratio of the system~\eqref{eq:linear}. The system is under-damped, thus the overshoot can cause constraint violations even when the steady-state $x_\infty$ lies in the constraint set $x_\infty\in\X$. We purposely chose a linear system with polytopic constraints~\eqref{eq:linear} since there are standard model-based methods~\cite{bertsekas1972} to compute the actual maximal admissible set $\Lambda_\infty$ for this problem to compare with our data-driven admissible set $\bar \Lambda$.

To create the data-set~\eqref{eq:data} needed for Algorithm~\ref{alg:mci}, we sampled the state trajectories of the system~\eqref{eq:linear} with sampling time of $\Delta t=0.1s$. For every command $\bar r_i=-1.2+0.02(i-1)\in[-1.2,~1.2]$, $i\in\naturals$, we generated $N_T=5$ state trajectories of length $T=400$ with random initial conditions. The state equilibrium $x_\infty$ was approximated by~\eqref{eq:xinf}. For the basis functions $\phi$, a set of thin-plate spline functions {$\phi_i(x)=\|x-c_i\|^2\ln{\|x-c_i\|}$, $i=1,\cdots,N_\phi$} were used with the centers $c_i$ located on a $14\times 14$ rectangular grid that envelope the state samples. 
{Commercial \ac{sdp} solvers encountered numerical issues attempting to solve~\eqref{eq:data-sdp}. In contrast, \texttt{MATLAB}'s \texttt{linprog} successfully solved the \ac{lp}~\eqref{eq:lp-alpha} for every $\bar r_i\in[-1.2,~1.2]$ where $\gamma=0.005$. However, the \ac{lp}~\eqref{eq:lp-alpha} was infeasible when $\gamma=0$ and $\gamma=0.1$. Satisfying~\eqref{eq:lp-decrease} for $\gamma=0$, i.e. Lyapunov decrease, is difficult for the samples close to the approximated equilibrium $x_\infty$. Conversely, $\gamma=0.1$ makes~\eqref{eq:lp-decrease} infeasible for the samples far from $x_\infty$. Furthermore, we observed that the size of $\bar \Lambda$ decreases as $\beta$ increases since it tightens the condition~\eqref{eq:lp-decrease}, making \ac{cg} more conservative.}

\begin{figure}[t]
\centering
\begin{subfigure}{0.237\textwidth}
\centering
    \includegraphics[width=\linewidth]{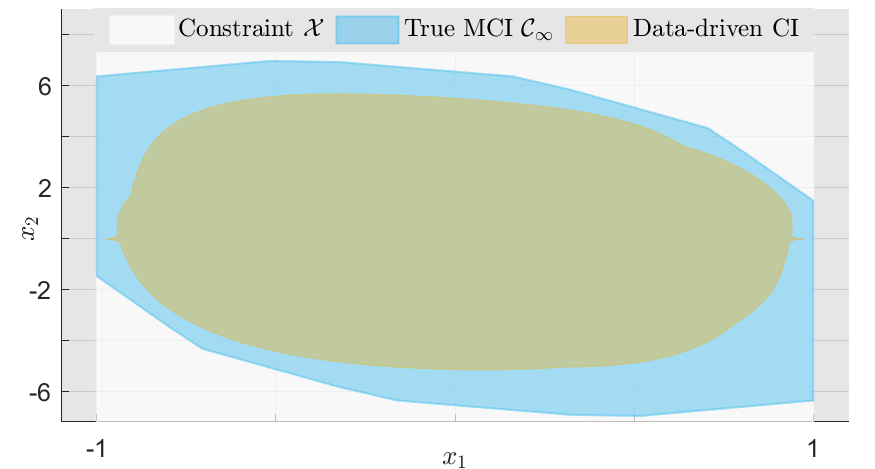}
    \caption{Top view}
    \label{fig:xx}
\end{subfigure}
\begin{subfigure}{0.237\textwidth}
\centering
    \includegraphics[width=\linewidth]{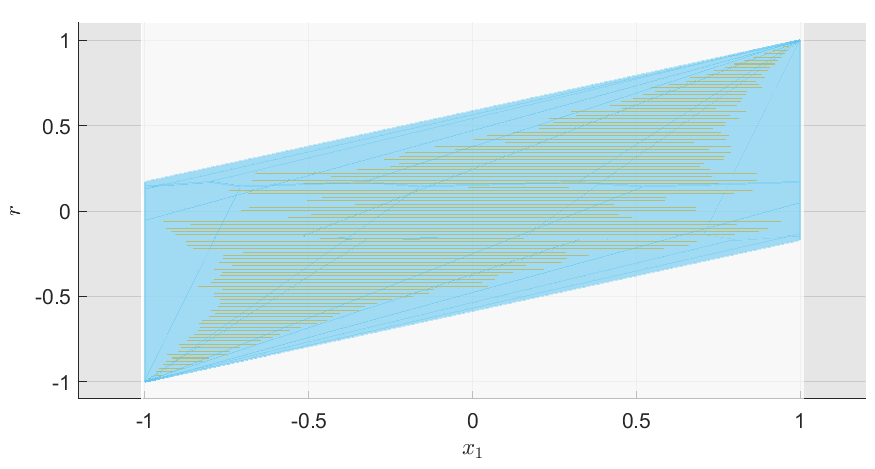}
    \caption{Front view}
    \label{fig:x1r}
\end{subfigure}%
\hfill
\begin{subfigure}{0.237\textwidth}
\centering
    \includegraphics[width=\linewidth]{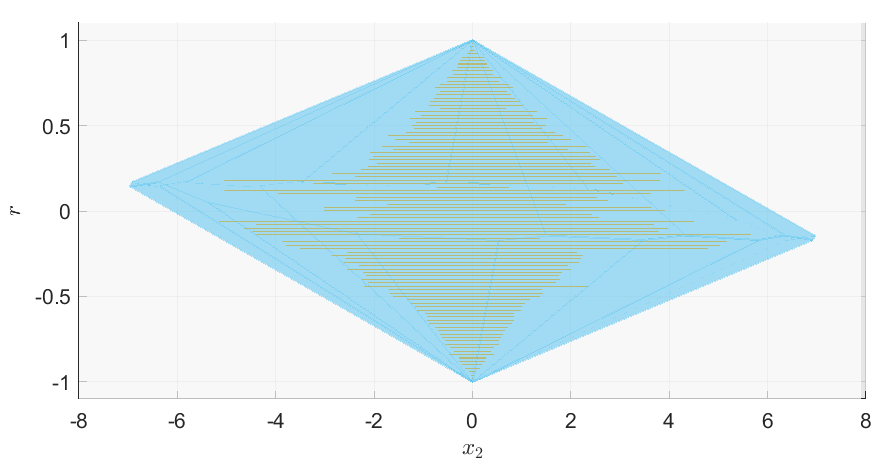}
    \caption{Side view}
    \label{fig:x2r}
\end{subfigure}
    \caption{The blue polyhedron is the maximal admissible set $\Lambda_\infty$ of the system~\eqref{eq:linear}, while the orange balloon-like set is our data driven admissible set $\bar \Lambda$, computed by Algorithm~\eqref{alg:mci}. The figures are the projection of the corresponding sets onto (a) $x_1, x_2$ (b) $x_1, r$, and (c) $x_2, r$ planes, where $x_1$, $x_2$, and $r$ are respectively position, velocity and command.}
    \label{fig:linear-mci}
\end{figure}

Fig.~\ref{fig:linear-mci} shows the comparison of the true $\Lambda_\infty$ and data-driven $\bar \Lambda$ admissible sets from different angles, where $\gamma=0.005$ and $\beta=0$. For instance, Fig~\ref{fig:linear-mci}.(a) shows the max-\ac{ci} set $\C_\infty$ and our data-driven \ac{ci} set~\eqref{eq:union}, which are the projection of the admissible sets onto the state plane.  

\begin{figure}[t]
\begin{subfigure}{0.475\textwidth}
\centering
    \includegraphics[width=\linewidth]{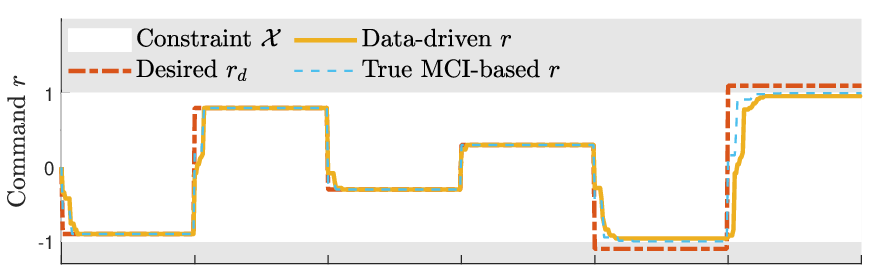}
    \vspace{-1.4em}
\end{subfigure}
\begin{subfigure}{0.475\textwidth}
\centering
    \includegraphics[width=\linewidth]{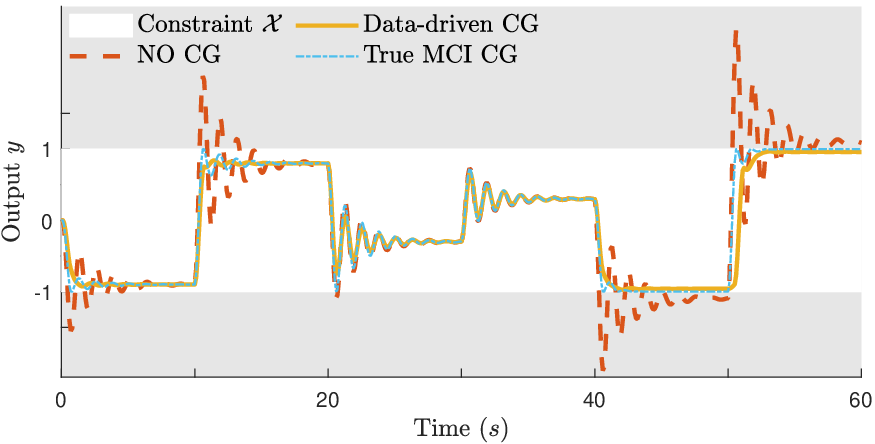}
\end{subfigure}
    \centering
    \caption{The top shows the desired command $r_d$, along with the modified commands from our data-driven {\ac{cg}} and the model-based max-\ac{ci} set \ac{cg}. The 
    bottom shows an output $y=[1,0]x$ of the linear system~\eqref{eq:linear} to these commands.}
    \label{fig:linear-output}
\end{figure}

The performance of our data-driven {\ac{cg}} is shown in Fig.~\ref{fig:linear-output} for controlling the output position $y=[1,0]x$ of the system~\eqref{eq:linear}. For this experiment, {we used the exact discretization} of the system~\eqref{eq:linear}. {We further implemented \ac{cg} using the true maximal admissible set $\Lambda_\infty$ for comparison. The results show that our data-driven \ac{cg} successfully enforces the constraint, while its performance is close to that of the \ac{cg} with the true maximal admissible set $\Lambda_\infty$.}

\subsection{Autonomous lane-keeping}

We demonstrate the capability of our data-driven {\ac{cg}} in handling nonlinear systems through an autonomous lane-keeping problem. The following kinematic bicycle model from~\cite{rajamani2011vehicle} is borrowed for the vehicle dynamics, 
\begin{subequations}
\label{eq:bicycle}
    \begin{align}
        \dot y&=v\sin(\theta+u), \qquad y\in[-2,2],\\
        \dot \theta&= \frac{v}{l}\sin(u),
    \end{align}
\end{subequations}
where the state $x=[y, \theta]^\trans$ includes the lateral position $y$ of the center of mass and the inertial heading angle $\theta$. The variable $v$ represents the speed, and $l=1.6m$ is the distance from the center of mass of the vehicle to the rear axle.
To facilitate plotting the admissible set, we will consider the angle of the velocity with respect to the longitudinal axis of the car as the control input $u$. The state constraint set $\X=\{x: [1, 0]x \in [-2,2]\}$ represents the allowable maneuver of the vehicle from the center of the two-lane road. Here, the widths of the vehicle and each lane are $1.6 \, \text{m}$ and $3.6 \, \text{m}$, respectively. We design a locally stabilizing controller by linearizing the nonlinear bicycle model~\eqref{eq:bicycle} at the origin and nominal velocity $v$,
\begin{equation}
    \dot x=\begin{bmatrix} 0 & v \\ 0 & 0\end{bmatrix} x + \begin{bmatrix} v \\ v/l\end{bmatrix} u.
\end{equation}

We used a \ac{lqr} with the weights $Q=\text{diag}([100, 10])$ and $R=1$ for state tracking $x\to[r, 0]^\trans$ at the nominal speed $v=20~\text{m/s} \approx 45~\text{mph}$. Although the closed-loop system is almost overdamped under the \ac{lqr} controller at the nominal speed, it exhibits underdamped behavior at higher speeds. Our goal is to ensure constraint enforcement both at the nominal speed and at higher speed of $v=27~\text{m/s} \approx 60~\text{mph}$ using the \ac{lqr} inner controller designed for the nominal speed.

{To generate the data, we simulated the kinetic bicycle model~\eqref{eq:bicycle} in closed-loop with the nominal \ac{lqr} controller using \texttt{MATLAB}'s \textsc{ode45} solver where the state was sampled with a period of $\Delta t = 0.1$ s.} For every command set-point $\bar{r}_i = -1.2 + 0.02(i-1)\bar{r} \in [-2.4, 2.4]$, $i \in \mathbb{N}$, we generated $N_T = 5$ trajectories of length $T = 100$ with random initial conditions. We set the thin-plate spline basis $\phi$ similar to the previous example and approximated $x_\infty$ by~\eqref{eq:xinf}. We used this data to find $P$ by solving the \ac{lp}~\eqref{eq:lp-alpha}, where $\gamma = 0.01$ and $\beta = 0.001$. 
{
The true maximal admissible set $\Lambda_\infty$ for the bicycle dynamics~\eqref{eq:bicycle} is not known. However, we estimated $\Lambda_\infty$ by creating a grid of $20\times 20$ over the box $x\in[-2,2]$ and $\theta\in[-\tfrac{\pi}{2},\tfrac{\pi}{2}]$ in the state space and checked whether the state trajectory starting at the center of each cell violates the constraints.
}

\begin{figure}[t]
    \centering
    \begin{subfigure}[b]{0.237\textwidth}
        \includegraphics[width=\linewidth]{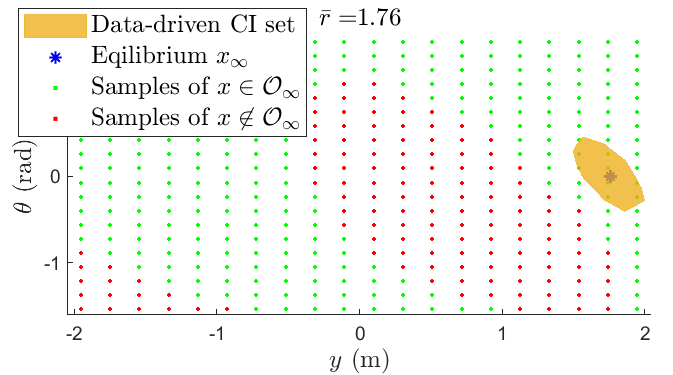}
    \end{subfigure}%
    \hfill
    \begin{subfigure}[b]{0.237\textwidth}
        \includegraphics[width=\linewidth]{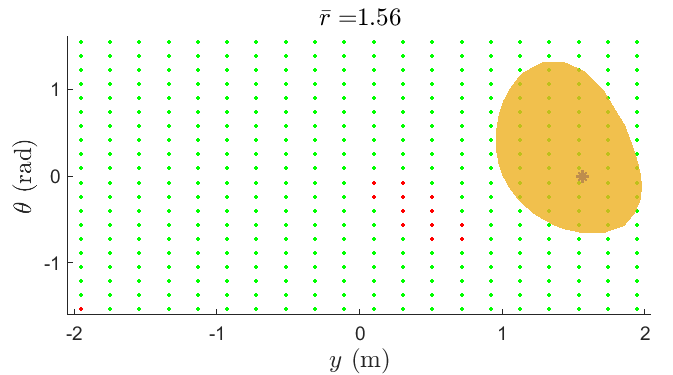}
    \end{subfigure}
    \begin{subfigure}[b]{0.237\textwidth}
        \includegraphics[width=\linewidth]{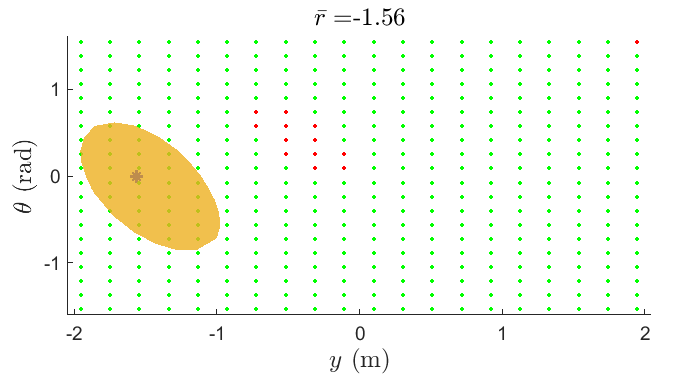}
    \end{subfigure}%
    \hfill
    \begin{subfigure}[b]{0.237\textwidth}
        \includegraphics[width=\linewidth]{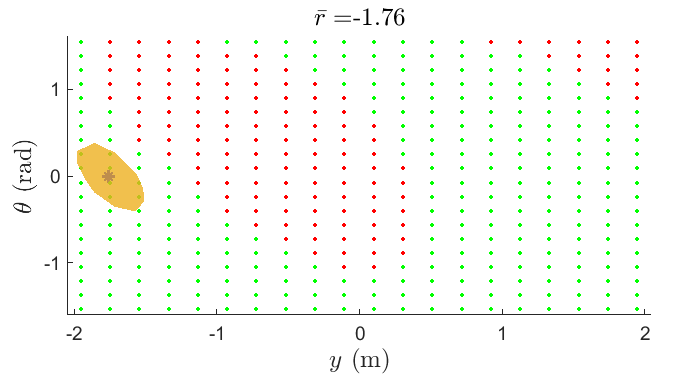}
    \end{subfigure}
    \caption{Admissible set for the nonlinear kinetic bicycle model~\eqref{eq:bicycle} generated by Algorithm~\ref{alg:mci}. Each slice is a \ac{pi} set $\O(\bar r)$ for the constant $\bar r$.}
    \label{fig:nonlinear-mci}
\end{figure}

\begin{figure}[t]
    \centering
    \begin{subfigure}[b]{0.475\textwidth}
        \includegraphics[width=\linewidth]{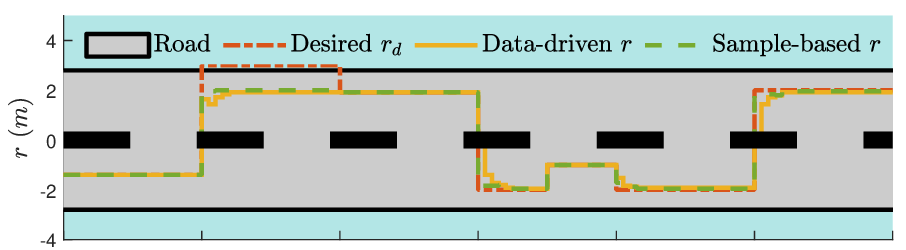}
    \end{subfigure}
    \begin{subfigure}[b]{0.475\textwidth}
        \includegraphics[width=\linewidth]{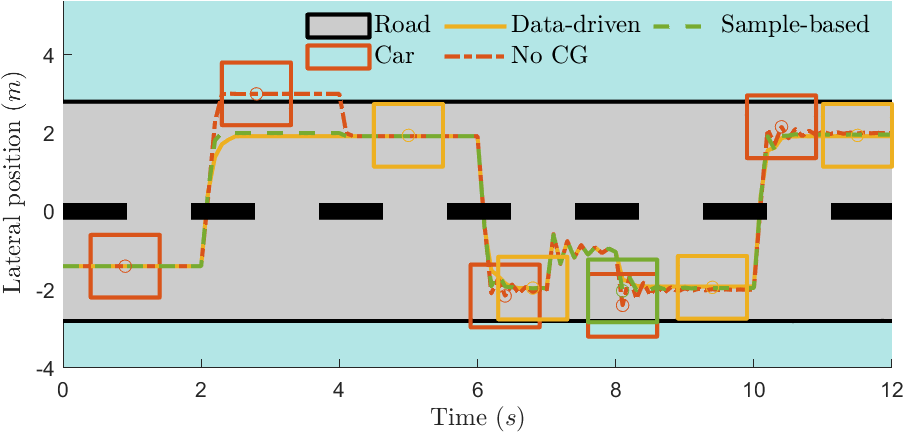}
    \end{subfigure}
    \caption{Constraint enforcement for the nonlinear kinetic bicycle model~\eqref{eq:bicycle} during overtaking maneuvers. The speed changes from $20~\text{m/s}$ to $27~\text{m/s}$ at $t=5$ s. The boxes indicate the position of the car at several instances.}
    \label{fig:nonlinear-output}
\end{figure}

{Fig.~\ref{fig:nonlinear-mci} shows the \ac{pi} sets computed by the sample-based and data-driven methods for four different $\bar{r}_i$. Although some of the data-driven learned \ac{pi} sets $\O(\bar r)$ are only a small subset of the constraint-satisfying samples, they encompass an area around the equilibrium, which provide acceptable performance of \ac{cg}. Fig.~\ref{fig:nonlinear-output} shows the command tracking of the system~\eqref{eq:bicycle} in closed-loop with the nominal \ac{lqr} controller {using \texttt{MATLAB}'s \textsc{ode45} solver and zero-order hold \ac{cg}}. We implemented \ac{cg} twice: once using the sample-based approach and another time using the data-driven admissible sets for comparison. We simulated several maneuvers, including overtaking maneuvers where the speed is changed from the nominal $20 \, \text{m/s}$ to a higher $27 \, \text{m/s}$ at $t=5 \, \text{s}$. Fig.~\ref{fig:nonlinear-output} shows that the car without \ac{cg} leaves the road several times, and a few times with the sample-based approach.}


\section{Conclusion}
\label{sec:conclusion}
Our data-driven method creates \ac{pi} sets directly from data for 
unmodeled systems. Additionally, we used these \ac{pi} sets to create \ac{ci} sets that solve control problems such as \acp{cg}. The computation of our \ac{pi} sets can be done offline, and it can be reduced to a \ac{lp}.
\vspace{-1em}

\bibliographystyle{IEEEtran} 
\bibliography{ref} 

\end{document}